# Terahertz and Infrared Studies of Antiferroelectric Phase Transition in Multiferroic Bi$_{0.85}$Nd$_{0.15}$FeO$_3$

V. Goian[1], S. Kamba[1*], S. Greicius[2], D. Nuzhnyy[1], S. Karimi,[3] and I. M. Reaney[3]

[1]Institute of Physics, Academy of Sciences of Czech Republic, Na Slovance 2, Prague 8, Czech Republic
[2]Faculty of Physics, Vilnius University, Vilnius, Lithuania
[3]Department of Engineering Materials, University of Sheffield, Sheffield, United Kingdom

**Abstract**

High-frequency dielectric studies of Bi$_{0.85}$Nd$_{0.15}$FeO$_3$ ceramics performed beteween 100 and 900 K reveal hardening of most polar phonons on cooling below antiferroelectric phase transition, which occurs near 600 K. Moreover, a strong THz dielectric relaxation is seen in paraelectric phase. Its relaxation frequency softens on cooling towards T$_C$ ≈ 600 K, its dielectric strength simultaneously decreases and finaly the relaxation disappears from the spectra below 450 K. Both phonon and dielectric relaxation behavior is responsible for a decrease in the dielectric permittivity at the antiferroelectric phase transition. Origin of unusual strong THz dielectric relaxation in paraelectric phase is discussed. Bi$_{0.85}$Nd$_{0.15}$FeO$_3$ structrure lies on the phase boundary between polar rhombohedral and non-polar orthorhombic phase and owing to this the polarization rotation and polarization extension can enhance the piezoelectric response of this system. Similarities and discrepancies with lead-based piezoelectric perovskites, exhibiting morphotrophic phase boundary between two ferroelectric phases, are discussed.

**Introduction**

Bismuth ferrite BiFeO$_3$ (BFO) belongs to a scarce class of multiferroic materials, exhibiting simultaneously ferroelectric (FE) and antiferromagnetic ordering above room temperature (see review [1]). Coupling between magnetic and FE order parameters opens the way to a wide variety of possible applications such as magnetic sensors, FeMRAMs, actuators or transducers. BFO is particularly attractive due to its high Curie (~1100 K) and Néel (~640 K) temperatures, therefore magnetoelectric devices based on BFO could operate at room and elevated temperatures. Moreover, BFO has exceptionally high FE polarization reaching almost 100 μC/cm$^2$ making this system promising for FeRAMs.

However, before any practical applications can take place, three major problems of BFO must be addressed: (1) frequently high porosity and multi-phase composition of ceramics and films, (2) high leakage current at room temperature, suppressing the superior FE polarization, and (3) spiral spin modulation[2] superimposed on the G-type antiferromagnetic (AFM) spin ordering, inhibiting remnant magnetization and the linear magnetoelectric effect.

There are a few possible strategies to overcome these limitations: careful sintering process, growth of epitaxial films, application of high magnetic field[3] or doping, the last being the most popular. Since substitution of B-site ions drastically decreases magnetic ordering temperature, scientists turned their attention to exchanging the A-site cations. Among many various A-site substituted BFO the most interesting structural and dielectric behavior was discovered in rare-earth substituted BFO. It was shown that if La, Nd, Sm, Gd, and Dy partially substitute for $Bi^{3+}$ in BFO, the structure dramatically changes.[4,5,6,7,8,9,10,11] As dopant concentration increases, the initial FE rhombohedral *R3c* structure (with 2 formula units (Z) per unit cell $\sqrt{2}a_c \times \sqrt{2}a_c \times a_c$, where $a_c \approx 4$ Å is the lattice parameter of the ideal cubic perovskite) transforms into an antiferroelectric (AFE) $PbZrO_3$-type orthorhombic phase with *Pbam* structure[11] (Z=8, unit cell is $\sqrt{2}\ a_c\ x\ 2\sqrt{2}\ a_c\ x\ 2a_c$) and then finally to a paraelectric (PE) orthorhombic phase (*Pbnm*, $\sqrt{2}a_c \times \sqrt{2}a_c \times 2a_c$, Z=4). The exact symmetry of the AFE phase is still subject to debate in the literature. The structure refines from X-ray and neutron diffraction data to the $PbZrO_3$ Pbam cell, but electron diffraction reveals weak superstructure of the type ¼{00l} which results in a larger unit cell ($\sqrt{2}\ a_c\ x\ 2\sqrt{2}\ a_c\ x\ 4a_c$, Z=16) and *Pbnm* symmetry.[11]

Despite the minor descrepancy, it is clear that all rare-earth substituted BFO samples exhibit generally similar phase diagrams with AFE morphotropic phase boundary (MPB) between FE and PE phase, where the critical concentrations depend on average ionic A-site radius.[7,8] The phase diagram of $Bi_{1-x}Nd_xFeO_3$ is shown in Fig. 1. Thanks to the phase boundary between polar and non-polar phases, the polarization rotation and polarization extension [12] is considered to enhance electromechanical coupling with some authors reporting piezoelectric coefficients 2 to 3 times greater at rather than away from the boundary.[8,13,14]

A-site substitutions have been shown to increase the Néel temperature ($T_N$), but there is descrepancy between the $T_N$ observed by neutron diffraction and the appearance of the peak in latent heat in differential scanning calorimetry. Irrespective of the Re dopant, AFM structure remains G-type, but structural phase transitions are accompanied by reorientation of magnetic dipoles within the G-type AFM structure resulting in a significant increase of magnetization.[9]

Dielectric properties (polarization, piezoelectric coeficients and complex permittivities ε*) of rare-earth substituted BFO were investigated mainly at room temperature.[8,10,13] Only permittivities of $Bi_{1-x}Nd_xFeO_3$ (BNFO) ceramics (x=0-0.2) were investigated at high-temperatures up to 400 ºC.[6] Dielectric permittivities measured at 100 kHz exhibit anomalies in the vicinity of structural phase transitions (PT), but the values of $\varepsilon^*$ are influenced by conductivity, which enhances both permittivities and dielectric losses due to Maxwell-Wagner polarization. Hence, measurements to date in the radio frequency range are considered unreliable at determing $T_C$ in these ceramics. The influence of conductivity, however, dramatically decreases with frequency and therefore high-frequency/high-temperature dielectric studies are desirable.

In this contribution we concentrate on THz transmission and infrared (IR) reflectivity studies of $Bi_{0.85}Nd_{0.15}FeO_3$ lying close to AFE/FE boundary. The obtained complex $\varepsilon^*$ spectra between 150 GHz and 20 THz are most likely not influenced by conductivity. The spectra allow to distinguish phonon contribution to $\varepsilon^*$, which exhibits a drop at $T_C$ on cooling due to hardening of phonon frequencies on lowering temperature. Simultaneously, a strong dielectric relaxation was revealed below 1 THz in paraelectric phase. Its relaxation frequency and dielectric strength decreases on cooling and finally the relaxation disappears below 450 K. The origin of the relaxation and its relation to the strong piezo-response is discussed.

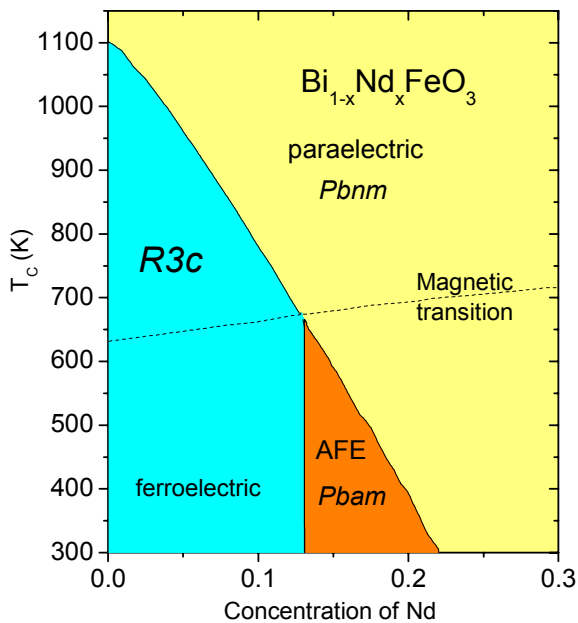

FIG. 1. Schematic structural and magnetic phase diagram of $Bi_{0.85}Nd_{0.15}FeO_3$ (plotted according to data in Ref. [11]). Ferroelectric $R3c$ phase has $Z=2$, paraelectric $Pbnm$ phase $Z=4$ and antiferroelectric $Pbam$ phase $Z=8$. Electron diffraction revealed even $Z=16$ and $Pbnm$ structure in AFE phase. Néel temperature of G-type antiferromagnetic phase increases with Nd concentration (dashed line).

**Experimental details**

Preparation of more than 95% dense ceramics of BNFO was described in detail elsewhere.[6,7] IR reflectivity spectra were taken from pellets polished on one-side (thickness ~ 1 mm) whereas THz transmission spectra were obtained from parallel ground pellets polished to a mirror finish on both sides (thickness ~ 100 μm).

IR reflectivity measurements were performed using a Fourier Transform IR Spectrometer Bruker IFS 113v in the temperature range of 100 K ÷ 900 K and frequency range between 15 and 3000 cm$^{-1}$ (0.45 - 90 THz) with the resolution of 2 cm$^{-1}$. A helium-cooled Si bolometer operating at 1.6 K was used as a detector for low-temperature measurements while a pyroelectric DTGS detector was used above room temperature. Time-domain THz spectroscopy measurements were performed in the range of 150 GHz and 2 THz. Linearly polarized THz probing pulses were generated using a Ti:sapphire femtosecond laser whose pulses iluminated an interdigitated photoconducting GaAs switch. THz signal was detected using the electro-optic sampling with 1 mm thick [110] ZnTe crystal. An Optistat CF Oxford Instruments continuous flow helium cryostat was used for cooling the samples down to 100 K and a commercial high temperature cell Specac P/N 5850 was used for heating the samples up to 900 K in both IR and THz spectrometers.

**Results**

Fig. 2 shows experimental IR and THz reflectivity spectra of BNFO ceramics in the range of phonons, i.e. below 650 cm$^{-1}$. The spectra at higher frequencies (up to 3000 cm$^{-1}$) were measured only at room temperatures, because they exhibit a frequency and temperature independent value determined by the high-frequency permittivity $\varepsilon_\infty$ = 7.3 stemming from the electron absorption processes in the visible-UV range.

The IR reflectivity spectra $R(\omega)$ are related to the complex permittivity $\varepsilon^*(\omega)$ by the formula[15]

$$R(\omega) = \left| \frac{\sqrt{\varepsilon^*(\omega)} - 1}{\sqrt{\varepsilon^*(\omega)} + 1} \right|^2 \qquad (1)$$

For the spectra fits a generalized-oscillator model was used with the factorized form of the complex permittivity[15]

$$\varepsilon^*(\omega) = \varepsilon'(\omega) - i\varepsilon''(\omega) = \varepsilon_\infty \prod_j \frac{\omega_{LOj}^2 - \omega^2 + i\omega\gamma_{LOj}}{\omega_{TOj}^2 - \omega^2 + i\omega\gamma_{TOj}} \qquad (2)$$

where $\omega_{TOj}$ and $\omega_{LOj}$ are transverse and longitudinal frequency of the j-th polar phonon, respectively, $\gamma_{TOj}$ and $\gamma_{LOj}$ are their damping constants and $\varepsilon_\infty$ denotes the high frequency permittivity resulting from electronic absorption processes. The IR reflectivity spectra are less accurate below 50 cm$^{-1}$; therefore the IR reflectivity spectra were fitted simultaneously with more reliable THz $\varepsilon^*(\omega)$ spectra. Complex IR permittivities and $\omega_{TOj}$ phonon frequencies obtained from the fits are shown in Figs. 3 and 4.

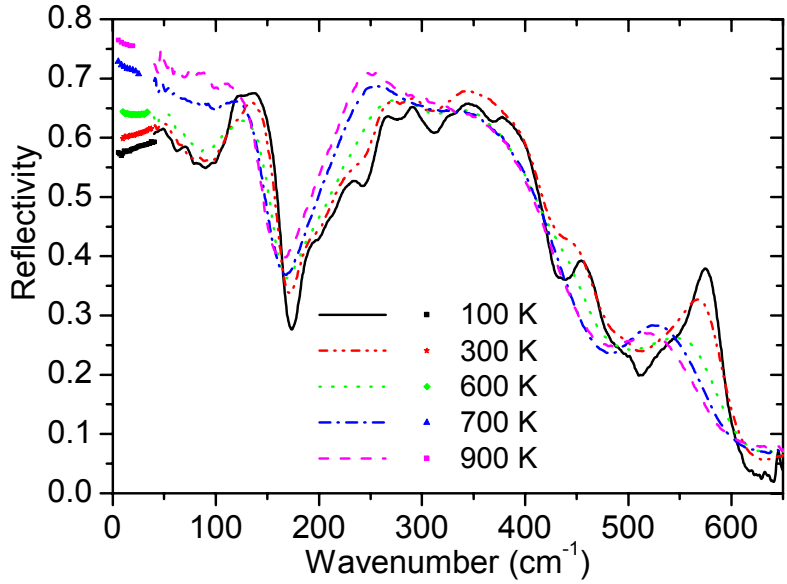

FIG. 2. Infrared reflectivity spectra at selected temperatures. Low-frequency spectra plotted below 50 cm$^{-1}$ were calculated from THz dielectric spectra using Eq. (1).

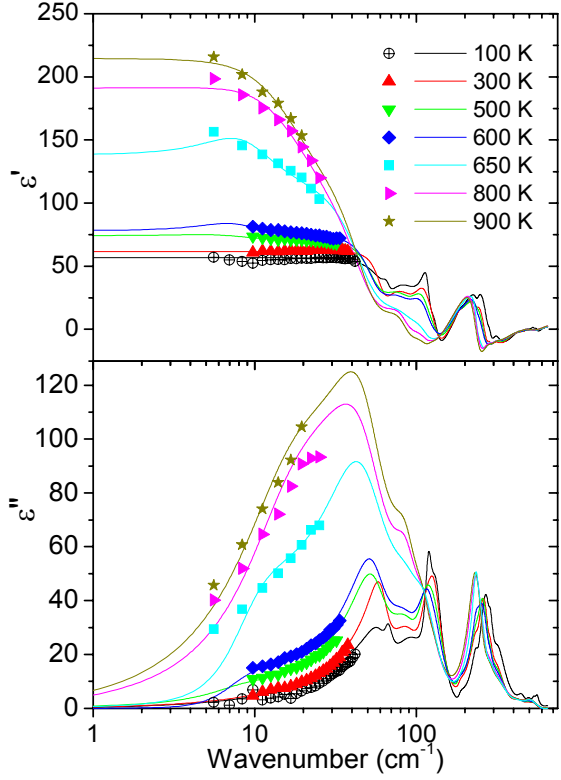

FIG. 3. Real and imaginary parts of complex dielectric permittivity obtained from the fits of IR reflectivities and THz $\varepsilon^*(\omega)$ spectra at selected temperatures. Symbols at low frequencies are experimental THz data.

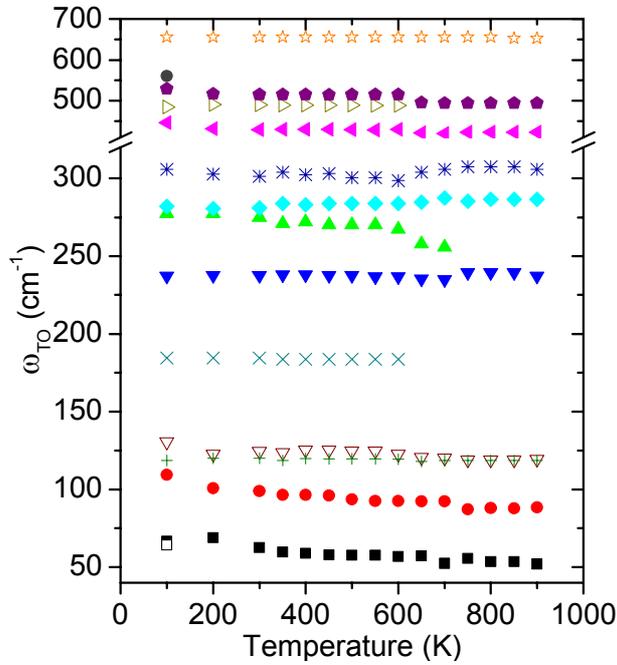

FIG. 4. Temperature dependence of transverse phonon frequencies obtained from the fits of THz and IR spectra. Appearance of several new phonons due to lowering of crystal symmetry below 600 K is seen.

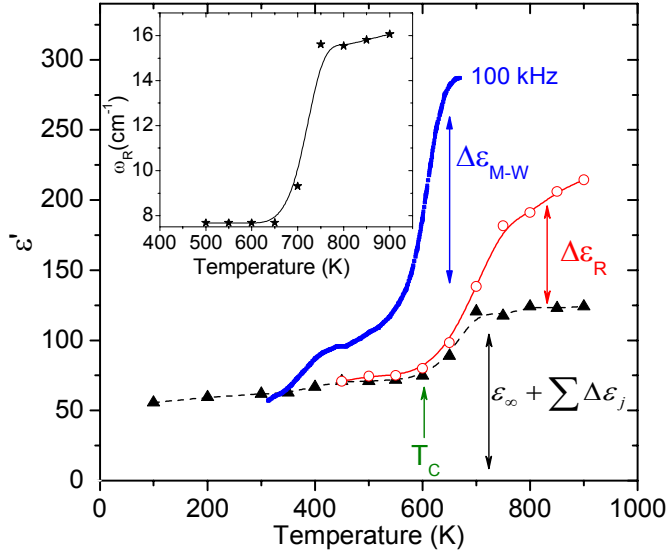

FIG. 5. Temperature dependence of static permittivity calculated from the fits of THz and IR spectra compared with published[6] 100 kHz ε' data. Sum of electronic and phonon contributions ($\varepsilon_\infty + \sum \Delta\varepsilon_j$) exhibits plateau between 700 and 900 K and drop down near $T_C$. Contribution $\Delta\varepsilon_R$ of dielectric relaxation diminishes on cooling and fully disappears from the spectra below 450 K. Experimental permittivity obtained at 100 kHz[6] exhibits the highest value due to contribution of Maxwell-Wagner polarization. Inset shows temperature dependence of relaxation frequency $\omega_R$ of the central (relaxational) mode.

The dielectric contribution $\Delta\varepsilon_j$ of the *j*-th mode to static permittivity may be calculated by[15]

$$\Delta\varepsilon_j = \frac{\varepsilon_\infty}{\omega_{TOj}^2} \frac{\prod_k (\omega_{LOk}^2 - \omega_{TOj}^2)}{\prod_{k \neq j} (\omega_{TOk}^2 - \omega_{TOj}^2)} \quad (3)$$

The sum of phonon and electronic contributions to the static permittivity may be written as

$$\varepsilon'(0) = \varepsilon_\infty + \sum_{j=1}^{n} \Delta\varepsilon_j \quad (4)$$

and its temperature dependence is plotted in Fig. 5. Above 450 K, a heavily damped excitation appears in the spectra with relaxation frequency $\omega_R$ = 8 cm$^{-1}$ and negligible dielectric strength $\Delta\varepsilon_R$=3. Above $T_c \approx$ 600 K the lowest frequency excitation becomes overdamped and dielectric contribution $\Delta\varepsilon_R$ of the relaxation dramatically increases on heating and reaches value about 90 at 900 K. For the fitting of the spectra we did not use Debye model of the relaxation

$$\varepsilon^* = \varepsilon_{\infty'} + \frac{\Delta\varepsilon_R}{1 + i\frac{\omega}{\omega_R}} \quad (5)$$

because this model does not satisfy the *f* sum rule[16]

$$\int_0^\infty \omega \varepsilon''(\omega) d\omega = const.  \quad (6)$$

The Debye relaxation model gives non-zero dielectric losses at all frequencies above $\omega_R$ and therefore $\int_0^\infty \omega \varepsilon''(\omega) d\omega = \infty$. Accordingly, the Debye relaxation spoils the shape of IR reflectivity spectra far above $\omega_R$. Moreover, it is not possible to combine this model with the phonon damped oscillator model in Eq. 2, which is in product form (additionally a sum of oscillators model was attempted instead of Eq. 2, but the fit of reflectivity was less accurate). For the fits only Eq. 2 was utilized and the relaxation was modeled by an oscillator with much higher damping than the eigenfrequencies. If $\gamma_{TO} = \gamma_{LO} > 2\omega_{TO}$ (overdamped oscillator), the relaxation frequency $\omega_R = \omega_{TO}^2 / \gamma_{TO}$ corresponds to frequency of maximum of dielectric loss. In our case $\gamma_{LO} > \gamma_{TO}$ and $\gamma_{TO} < 2\omega_{TO}$ therefore $\omega_R \neq \omega_{TO}^2 / \gamma_{TO}$. We determined $\omega_R$ from the frequency of maxima of dielectric loss $\varepsilon''(\omega)$ spectra, which makes physical sense. The dielectric relaxation has a strong influence on the shape of the IR reflectivity above $T_C$. It dramatically enhances the value of reflectivity at low frequencies, causing the disappearance of the reflectivity minimum near 90 cm$^{-1}$ (see Fig. 2) and remarkably enhances the THz permittivity and losses above $T_C$ (see Fig. 3). Temperature dependence of the relaxation frequency is plotted in the inset of Fig. 5.

**Discussion**

Displacive ferroelectric PT's are driven by a polar (i.e. IR active) optical soft mode with frequency $\omega_{SM}$, which softens (i.e. reduces frequency) on cooling towards $T_C$ according to the Cochran law $\omega_{SM}^2 = A(T - T_C)$ and hardens on further cooling below $T_C$.[17] Due to conservation of the oscillator strength $f_{SM}$ ($f_{SM} = \Delta\varepsilon_{SM}\omega_{SM}^2 = const$) in the case of uncoupled phonons), static permittivity exhibits the Curie-Weiss maximum at $T_C$. More frequently, order-disorder phase transitions are observed, in which the soft mode is not an optical phonon, but a dielectric relaxation.[18,19] In these cases some ions in the paraelectric phase are dynamically disordered among equivalent positions (as exemplified by Ti and Nb disorder in BaTiO$_3$ and KNbO$_3$, respectively)[20] and the relaxation is a manifestation of hopping of B-site cations among equivalent positions. Usually, the relaxation frequency $\omega_R$ linearly decreases on cooling towards $T_C$ according to a modified Cochran law $\omega_R = A(T - T_C)$. The strength of the dielectric relaxation defined as $f_R = \Delta\varepsilon_R \omega_R$ is usually also temperature independent above $T_C$ and therefore permittivity increases on cooling towards $T_C$. A very commonly observed case is the so-called 'crossover' of displacive and order-disorder behavior of the PT.[18,21] In this case the PT is far above $T_C$ driven by the optical soft mode. Close to $T_C$ the soft mode frequency saturates and a soft dielectric relaxation appears (i.e. the central mode), which receives an oscillator strength from the coupled soft polar phonon and finally drives the PT. Below $T_C$ the ions are usually ordered, i.e. the hopping of ions disappears and therefore the strength $f_R$ dramatically decreases on cooling (actually, $f_R$ transfers to phonons due to $f$ sum rule) and finally the relaxation expressing dynamical disorder of ions disappears from the spectra. Simultaneously, another dielectric relaxation related to dynamics of FE domain walls frequently appears below $T_C$.

AFE as well as improper-ferroelectric PT's are driven by a lattice instability at the Brillouin zone boundary, which causes multiplication of unit cell below the PT.[22] IR and Raman spectroscopic techniques can study only phonons from Brillouin zone centre due to conservation of $q$ wavevectors of phonons and photons. Therefore, the soft mode cannot be IR or Raman active above $T_C$ and it may become IR and Raman active only below $T_C$ due to folding of the Brillouin zone in the low-temperature phase. As the consequence of multiplication of unit cell below AFE or improper-ferroelectric PT, IR selection rules change and many new modes arise in the spectra below $T_C$.

In the case of BNFO, rather small change of number of polar phonons with temperature is observed (see Fig. 4). Only five new polar modes appear below $T_C$ due to multiplication of the unit cell (see Table I with the phonon parameters obtained from the IR spectra fits at 100 and 900 K). Mainly the new mode (No. 8 in Table I), which appears at 700 K near 250 cm$^{-1}$, exhibits a remarkable hardening to 277 cm$^{-1}$ on cooling to 100 K. Also the modes No. 1, 3, 11 and 13 exhibit significant hardening from 52, 88, 422 and 490 cm$^{-1}$ (at 900 K) to 64, 109, 446 and 530 cm$^{-1}$ (100 K), respectively (see Table I and Fig. 4). The hardening of the all phonons is responsible for decrease of their dielectric strengths and therefore also the sum of phonon permittivity decreases below $T_C$ in Fig. 5. The most remarkable influence on the shape of IR and THz spectra has the relaxation mode, which weakens on cooling and finally disappears from the spectra below ~450 K. Its relaxation strength transfers to the phonons, but in spite of it the phonon contribution to $\varepsilon^*$ drops down below $T_C$ because of phonon hardening.

In other antiferroelectrics like $PbZrO_3$ a peak in permittivity at $T_C$ is usually seen and then a sharp drop-down below $T_C$ due to the first order PT. In our case the phonon contribution to $\varepsilon^*$ in BNFO exhibits temperature independent $\varepsilon'$ in paraelectric phase and jump down at $T_C$. Even stronger decrease of $\varepsilon'$ exhibits contribution of dielectric relaxation, but its dielectric strength does not exhibits any saturation or peak above $T_C$, its value continuously increases on heating above $T_C$. This unusual behavior will be discussed lower.

In $PbZrO_3$, the dielectric anomaly is caused mainly by a dielectric relaxation (central mode) below the phonon frequencies, but $\varepsilon'(T)$ exhibits a peak (of ~3500 near $T_C$) in $PbZrO_3$ ceramics and single crystal.[23,24] The contribution of central mode $\Delta\varepsilon_R$ in $PbZrO_3$ can be drastically reduced if the ceramics exhibit cracks in which situation the peak in $\varepsilon'(T)$ disappears, permittivity is four-times smaller and only a sharp step down occurs in $\varepsilon'(T)$ at $T_C$.[24] The small peak in $\varepsilon'(T)$ and the sharp drop down due to the first-order PT was observed also in hydrogen-bonded antiferroelectrics like $NH_4H_2PO_4$ and $CsH_2AsO_4$.[25,26] Our observed continuous increase of static $\varepsilon'$ above $T_C$ is similar to the dielectric behavior of the proton-bonded $KD_2(SeO_3)_2$ [27] and $TlH_2PO_4$,[28] but the absolute values of $\varepsilon'$ are much smaller in these materials than in BNFO. It is worth noting that the shape of our temperature dependence of static $\varepsilon'$ calculated from THz and IR spectra qualitatively corresponds to experimental 100 kHz data published by Karimi et al.[6] (see Fig. 5), but the latter values are higher due to contribution of Maxwell-Wagner polarization stemming from non-zero conductivity of the sample.

The shape of low-temperature IR reflectivity spectra of BNFO reminds our spectra of pure BFO ceramics published several years ago[15], although the crystal structures of both samples are different. In Table II the factor group analysis is presented for all possible crystal structures of BNFO (Fig. 1) using the tables from Ref.[29]. One can see that only 13 polar modes are expected in the FE *R3c* structure of pure BFO, while 45 or even 109 IR active modes are allowed for AFE *Pbam* or possible *Pbnm* phase with Z=8 or Z=16 (possible structures are discussed in the Introduction). We have observed only 15 polar phonons in our low-temperature IR spectra (Fig. 4). The probable reason is that the newly activated phonons are weak and have frequencies close to the modes of the FE *R3c* structure, so that the modes become overlapped and cannot be resolved from the IR spectra. The possible splitting of the modes has influence on effectively much higher phonon damping observed in BNFO IR reflectivity spectra than in pure BFO.[15] The high-temperature IR reflectivity of BNFO differs from that of BFO mainly at low frequencies below 150 cm$^{-1}$ due to the presence of THz dielectric relaxation. This causes disappearance of the reflectivity minimum near 90 cm$^{-1}$ in BNFO above 700 K, while in BFO the minimum is present up to the highest investigated temperature of 950 K.[15] Static permittivity of BFO slightly and linearly increases on heating due to small phonon softening with rising temperature, which is typical for improper-ferroelectric phase transition in BFO.[15] In BNFO the phonon permittivity exhibits jump up at $T_C$ and plateau above it. Nevertheless, contribution of dielectric relaxation grows on heating above $T_C$ and therefore the static permittivity from the fit of THz and IR spectra also continuously increases.

Only 10 polar phonons were used for the fits of reflectivity spectra in PE phase, although 25 IR active modes are allowed in the PE phase (see Table II). Again, discrepancy between the number of predicted and observed polar phonons is caused mainly by overlapping of the phonons due to high phonon dampings at high temperatures. As a result it is impossible to distinguish all the allowed IR active phonons in the spectra.

Temperature dependence of the dielectric relaxation in paraelectric phase is not typical for order-disorder phase transitions. Relaxation frequency $\omega_R$ does not follow the modified Cochran law, i.e. does not soften linearly on cooling towards $T_C$ (see Fig. 5). Relaxation strength $f_R$ is not temperature independent, but strongly decreases on cooling. This causes absence of the peak in $\varepsilon'(T)$ at $T_C$. The PT also does not look like the first order one, because the decrease of $\varepsilon'$ on cooling is not step-wise at $T_C$, but rather smooth. Broadening of the phase transition can be caused by an inhomogeneous concentration of Nd in BNFO ceramic.

FE polarization in pure BFO is caused by lone pairs of electrons and displacements of Bi cations at the perovskite A-sites.[1] Simultaneously, the polar displacements of Bi are coupled to the $a^-a^-a^-$-type of oxygen octahedral tilting (Glazer notation[30]). In the PE phase the A-site cations are displaced in an antiparallel and coupled to the $a^-a^-c^+$ tilting. $T_C$ steeply decreases with Nd concentration in BNFO (Fig. 1) and the intermediate AFE phase ($a^-b^-c^-/a^-b^-c^+$) appears in a narrow concentration range probably due to frustration between the FE and PE phase.[31] In other words, the driving force for polar Bi displacements becomes weaker, which results in an antipolar order.

Very strong influence on the IR spectra and static permittivity has the THz dielectric relaxation. Where is it coming from? There are several possible explanations: 1) The relaxation expresses highly anharmonic (i.e. damped) rotations and tilting of oxygen octahedra in paraelectric phase. Below $T_C$ the relaxation disappears because the structure orders. 2) The relaxational mode is vibrations of Bi cations, which may become highly anharmonic due to a partial substitution by Nd cations. This causes dynamical disorder of A-site cations in the PE phase. Hopping of Bi and Nd cations among equivalent positions in PE phase may cause the THz relaxation. Below $T_C$ the A-site cations order and therefore the relaxation disappears from the spectra and ε' saturates at a lower value determined by the sum of phonon contributions (Fig. 5). Nevertheless, it is worth noting that the driving force of the AFE phase transition is a phonon from the Brillouin zone boundary, which causes the multiplication of unit cell below $T_C$ and the THz dielectric relaxation seen in our spectra doesn't drive the AFE PT, it expresses just some anharmonic vibration of Bi (Nd) cations in the PE phase. 3) The THz relaxation expresses hopping conductivity of oxygen vacancies and increase of its strength just at $T_C$ is only accidental. The final decision which of the above three mechanism is correct needs additional experiments on non-conducting samples.

The best piezoelectric properties are known in lead-based solid solutions like $Pb(Zr_xTi_{1-x})O_3$, PMN-PT, PZN-PT etc. All these systems exhibit a morphotropic phase boundary (MPB) and a frustrated intermediate phase on MPB between two FE phases. Thanks to very similar free energy in all the phases existing around the MPB, the polarization can easily rotate in an anisotropically flattened free energy profile and enhanced electromechanical response is observed. A common feature of all such systems is the existence of a strong dielectric relaxation seen in the THz frequency region close to $T_C$, which splits into two components below $T_C$.[32,33] One component stays in the range of $10^{10} - 10^{11}$ Hz down to lowest temperatures and the second moves to a lower-frequency region (Hz - MHz) and simultaneously broadens on cooling below $T_C$. The latter probably corresponds to FE domain wall motion, but the former one expresses the highly anharmonic vibrations of Pb cations related to polarization rotation at the MPB.

In the case of BNFO the AFE phase boundary exists between FE and PE phase. The free energy is also flattened near the phase boundary and enhanced electromechanical response may occur not only due to polarization rotation, but also mainly due to so called polarization extension.[12] In contrast to the previous case, the THz relaxation is observed only above $T_C$ in BNFO, while below $T_C$ only phonons contribute to ε'. It means that the polarization rotation and extension probably contribute less to electromechanical response than in lead-based piezoelectrics. From that reason the piezoelectric constants in BNFO are only slightly higher than in pure BFO.[13] On the other hand, more than threefold increase in the piezoresponse was observed in La, Dy or Sm substituted BFO.[8,13,14] In these materials the polarization rotation and extension probably enhance more the electromechanical response and it would be interesting to check the broad-band dielectric response whether these mechanism will be manifested in the spectra by some THz dielectric relaxation.

Finally, it can be concluded that the permittivity of Nd substituted BFO exhibits different temperature behavior than the pure BFO. In PE phase, the permittivity is much higher than in BFO due to presence of the THz dielectric relaxation probably describing strongly anharmonic vibration of Bi and Nd. The relaxation frequency softens on cooling towards the AFE PT ($T_C \approx 600$ K) and in contrast to other materials with order-disorder PT its relaxational strength $f_R$ continuously decreases and finally the relaxation disappears from the spectra in the AFE phase. Due to this behavior the permittivity exhibits no maximum at $T_C$. It gradually decreases on cooling and finally saturates below $T_C$. Phonon contribution to permittivity exhibits temperature independent value in paraelectric phase and jump down at $T_C$. Factor group analysis of phonons in all the crystal phases of BNFO was performed. Most of the allowed polar phonons were not resolved in IR reflectivity spectra due to high phonon damping and overlapping with other phonons. The low-temperature reflectivity spectrum of BNFO reminds the spectrum of BFO, but the modes are more damped due to Nd substitution. The possible influence of polarization rotation and extension to enhanced electromechanical properties of rare-earth substituted BFO (with composition near the AFE/FE phase boundary) and its relation to observed THz dielectric relaxation, is discussed.

**Acknowledgments**

This work was supported by the Czech Science Foundation (Project 202/09/0682) and AVOZ 10100520. In addition, the contribution of Ph.D. student V. Goian has been supported by projects 202/09/H0041 and SVV-2011-263303. The authors thank J. Petzelt for valuable discussions and critical reading of the manuscript.

TABLE I. Mode parameters obtained from the fits at 100 and 900 K. CM means central mode.

| No. | 100K | | | | | 900K | | | | |
|---|---|---|---|---|---|---|---|---|---|---|
| | $\omega_{TO}$ (cm$^{-1}$) | $\gamma_{TO}$ (cm$^{-1}$) | $\omega_{LO}$ (cm$^{-1}$) | $\gamma_{LO}$ (cm$^{-1}$) | $\Delta\varepsilon$ | $\omega_{TO}$ (cm$^{-1}$) | $\gamma_{TO}$ (cm$^{-1}$) | $\omega_{LO}$ (cm$^{-1}$) | $\gamma_{LO}$ (cm$^{-1}$) | $\Delta\varepsilon$ |
| CM | | | | | | 16 | 27 | 20.1 | 37.1 | 90.2 |
| 1 | 64 | 23.5 | 68.2 | 12.5 | 1.1 | 52 | 51.1 | 71.2 | 51.4 | 69 |
| 2 | 66.6 | 9.7 | 64.3 | 15.5 | 2.8 | | | | | |
| 3 | 109.4 | 88.6 | 109.6 | 36.8 | 0.6 | 88.3 | 34.2 | 94.6 | 39.5 | 7.8 |
| 4 | 118.6 | 14.8 | 166.6 | 17.6 | 13.3 | 119.5 | 79.2 | 119 | 16.4 | 10.3 |
| 5 | 130.5 | 26.4 | 123.6 | 17.6 | 13.1 | 118.6 | 17.1 | 156.1 | 32.7 | 9.5 |
| 6 | 184.5 | 18.0 | 185.0 | 21.9 | 0.08 | | | | | |
| 7 | 237.1 | 20.3 | 281.1 | 15.1 | 0.4 | 237.2 | 61.0 | 273 | 49.1 | 14.4 |
| 8 | 277.5 | 77.4 | 238.1 | 17.2 | 12.6 | | | | | |
| 9 | 282.0 | 17.0 | 304.0 | 34. | 3.0 | 286.6 | 62.2 | 304.0 | 62.9 | 3.8 |
| 10 | 305.8 | 56.6 | 424.0 | 50.8 | 0.9 | 305.8 | 105.5 | 403.8 | 94.3 | 0.6 |
| 11 | 446.0 | 68.2 | 476.2 | 32.3 | 0.6 | 422.2 | 131.9 | 458.8 | 70.0 | 0.5 |
| 12 | 484.4 | 53.7 | 522.0 | 34.4 | 0.2 | | | | | |
| 13 | 529.3 | 25.7 | 544 | 46.7 | 0.06 | 493.6 | 125.0 | 580.6 | 37.4 | 0.8 |
| 14 | 560.4 | 31.5 | 599 | 33.3 | 0.2 | | | | | |
| 15 | 655.7 | 122.0 | 690.5 | 131.4 | 0.2 | 652.0 | 122.0 | 690.5 | 147.0 | 0.2 |

TABLE II. Factor group analysis of all vibrational modes (including acoustic) in various crystal phases of $Bi_{1-x}Nd_xFeO_3$ shown in phase diagram in Fig. 1. Activities of the modes in various kinds of spectra are shown in brackets. $x,y,z$ mean IR active modes active in polarized $\mathbf{E}\|x$, $\mathbf{E}\|y$ and $\mathbf{E}\|z$ spectra, respectively. (-) means silent modes, the remaining marks mean Raman active modes.

| FE: $R3c$ ($C_{3v}^6$) | PE: $Pbnm$ ($D_{2h}^{16}$) | AFE: $Pbam$ ($D_{2h}^9$) | AFE: $Pbnm$ ($D_{2h}^{16}$) |
|---|---|---|---|
| $\sqrt{2}a_c \times \sqrt{2}a_c \times a_c$ | $\sqrt{2}a_c \times \sqrt{2}a_c \times 2a_c$ | $\sqrt{2}a_c \times 2\sqrt{2}a_c \times 2a_c$ | $\sqrt{2}a_c \times 2\sqrt{2}a_c \times 4a_c$ |
| Z=2 | Z=4 | Z=8 | Z=16 |
| 5 $A_1$ ($z,x^2+y^2,z^2$) | 10 $B_{1u}$ ($z$) | 12 $B_{1u}$ ($z$) | 40 $B_{1u}$ ($z$) |
| | 7 $A_g$ ($x^2, y^2, z^2$) | 16 $A_g$ ($x^2, y^2, z^2$) | 28 $A_g$ ($x^2, y^2, z^2$) |
| 10 $E$ ($x,y,x^2-y^2, xy,xz,yz$) | 8 $B_{2u}$ ($y$) | 18 $B_{2u}$ ($y$) | 32 $B_{2u}$ ($y$) |
| | 10 $B_{3u}$ ($x$) | 18 $B_{3u}$ ($x$) | 40 $B_{3u}$ ($x$) |
| | 5 $B_{1g}$ ($xy$) | 16 $B_{1g}$ ($xy$) | 20 $B_{1g}$ ($xy$) |
| | 7 $B_{2g}$ ($xz$) | 14 $B_{2g}$ ($xz$) | 28 $B_{2g}$ ($xz$) |
| | 5 $B_{3g}$ ($yz$) | 14 $B_{3g}$ ($yz$) | 20 $B_{3g}$ ($yz$) |
| 5 $A_2$ (-) | 8 $A_u$ (-) | 12 $A_u$ (-) | 32 $A_u$ (-) |

*Corresponding author: kamba@fzu.cz